\documentclass[prb,twocolumn,showpacs]{revtex4-1}
\usepackage{graphicx}
\usepackage{times}
\usepackage{color}
\input pdfcolor.tex
\usepackage{graphicx}
\usepackage{epsfig,color}
\usepackage{amsfonts,amssymb}
\usepackage{amsmath}
\usepackage{color}
\usepackage{epstopdf}
\definecolor{refcolor}{rgb}{1.0,0.0,0.0}

\newcommand{\be}{\begin{equation}}
\newcommand{\ee}{\end{equation}}   
\newcommand{\bea}{\begin{eqnarray}}
\newcommand{\eea}{\end{eqnarray}}
\newcommand{\ba}{\begin{array}}
\newcommand{\ea}{\end{array}}

\newcommand{\phrl}[1]{Phys.~Rev.~Lett., {\bf #1}}
\newcommand{\phrb}[1]{Phys.~Rev.~B, {\bf #1}}

\renewcommand{\k}{{\bf k}}

\begin{document}
\title{Highly anisotropic quasiparticle interference patterns in 
the spin-density wave state\\ of the iron pnictides}

\date{\today}

\author{Dheeraj Kumar Singh and Pinaki Majumdar}

\affiliation{
Harish-Chandra Research Institute,
  Chhatnag Road, Jhunsi, Allahabad 211019, India\\
\& Homi Bhabha National Institute, Training School Complex,
Anushakti Nagar, Mumbai 400085, India
}

\begin{abstract}
We investigate the impurity scattering induced quasiparticle interference 
in the ($\pi, 0$) spin-density wave phase of the iron pnictides. We use a
five orbital tight binding model and our mean field theory in the clean
limit captures key features of the Fermi surface observed in 
angle-resolved photoemission. We use a t-matrix formalism 
to incorporate the effect of doping induced impurities on this state.
The impurities lead to a spatial modulation of the local density of states 
about the impurity site, with a periodicity of $\sim 8a_{{\rm Fe}-{\rm Fe}}$ 
along the antiferromagnetic direction. The associated momentum space 
quasiparticle interference pattern is anisotropic, with major peaks located 
at $\sim (\pm \pi/4,0)$, consistent with spectroscopic imaging scanning 
tunneling microscopy. We trace the origin of this pattern to an elliptical 
contour of constant energy around momentum (0,0), with major axis oriented 
along the (0,1) direction, in the mean field electronic structure.
\end{abstract}


\maketitle

\section{Introduction}

The highly anisotropic electronic properties of the
iron pnictides, with broken four-fold rotation symmetry, 
have been a subject of intense 
research in recent times. Observed in angle-resolved 
photoemission spectroscopy \cite{yi,shimojima} (ARPES), 
neutron magnetic resonance \cite{fu} (NMR), 
and transport properties \cite{chu,tanatar}, such 
anisotropy is seen both in the low temperature 
collinear antiferromagnetic (AF) 
state and the high temperature,
unordered, `nematic' phase \cite{blomberg}. 

ARPES  reveal a significant 
energy splitting between the $d_{xz}$ and 
$d_{yz}$ orbitals below the 
tetragonal-to-orthorhombic transition - which may 
precede \cite{nandi} or coincide with \cite{rotter}
the spin density wave (SDW) transition.
The spin dynamics shows a strong two-fold 
anisotropy inside the orthorhombic domains 
that are formed below the structural 
phase transition \cite{fu}, optical spectra displays a 
significant in-plane anisotropy \cite{nakajima} 
upto photonic energies $\sim 2$eV, and
transport measurements 
show a larger conductivity in the 
antiferromagnetic direction 
compared to the ferromagnetic
direction. 

Spectroscopic imaging - scanning tunneling microscopy 
(SI-STM)
\cite{capriotti,sykora,kreisel,zhang,yamakawa,hirschfeld} 
provides insight into
the anisotropic electronic state.  
Quasiparticle interference 
(QPI) probed by SI-STM measures 
the modulation of the local density of states (LDOS)
induced by the impurity atoms.
QPI patterns in the metallic ($\pi,0$)
SDW state consist mainly of a quasi-one
dimensional feature extended along
the $q_x = 0$ line with a
weaker parallel feature
\cite{chuang,rosenthal,zhou}
at a distance  $\sim \pi/4$.
Such highly anisotropic features
have been attributed
to impurity induced states    
on the anisotropic magnetic background
\cite{allan,gastiasoro}.

The QPI probes the response of the ordered state
to a strong localized perturbation and several
attempts have been made to explain it.   
A reasonable description
of the ARPES and QPI data imposes constraints on the 
electronic theory of the reference state. 
Broadly three 
frameworks have been used to to model the QPI, each with some
limitation. 

(i)~In an effective band approach \cite{knolle} 
LDOS modulation is strongest along the
{\it ferromagnetic} direction while experimentally
it is in the AF direction. Corresponding
contours of constant energy (CCE) consist mainly of a
circular pocket around $\Gamma$, smaller pockets located inside,
and the electron pocket around Y. 
(ii)~A five-orbital model    
\cite{graser} used to study QPI either produces
patterns without a clear modulation \cite{plonka} 
or shows modulation \cite{zhang1} at an energy 
$\omega \sim -150$meV, much 
larger than in the experiments. In one of the studies, details of 
the reconstructed FSs are not provided, while in another one large FSs
consists of parallel running structures extending 
from $\Gamma$ to the zone boundary near X \cite{yi1,wang}. 
(iii)~First principles calculations \cite{mazin} indicate 
QPI peaks at (0, $\pm\pi$/4) and therefore the correct wavelength of
modulation, $\sim 8a_{{\rm Fe}-{\rm Fe}}$, but again along the
{\it ferromagnetic} direction. In this case, FSs consist of 
crescent like structure around $\Gamma$ with the broader 
part facing Y. So, either the 
wavelength, or the orientation, or the
energy of the QPI modulations remain inconsistent with
experiments.

In this paper, we report on the 
QPI in the ($\pi, 0$)-SDW state 
of an electron-doped 
iron pnictide. We  use mean field theory on 
a five-orbital tight-binding model to describe the
ordered state and a t-matrix calculation to
quantify (single) impurity effects. We find the
following:
(i)~Our mean field bands  
have several features
consistent with the ARPES measurements, 
{\it e.g}, a large elliptical pocket
around $\Gamma$ and adjacent four smaller pockets. 
(ii)~The  QPI is highly anisotropic, consisting of  
quasi-one dimensional 
peak structures running nearly along $q_x = \pm\pi/4$. 
(iii)~The real-space 
features consist of LDOS modulation with periodicity 
$\sim 8a_{{\rm Fe}-{\rm Fe}}$ along the antiferromagnetic 
direction as observed in the STM measurements.  
The period of modulation along the AF direction is 
robust against change 
in the quasiparticle energy though the strongest 
modulation can shift to other direction.

\section{Model and method}

In order to study QPI in the SDW state, we consider a
 five-orbital tight-binding Hamiltonian defined in
the Fe-As planes, the kinetic part of which is given by
\begin{equation} 
{H}_0 = \sum_{\k} \sum_{\mu,\nu} \sum_{\sigma} 
\varepsilon^{\mu\nu}_{\bf k} 
d_{{\bf k}\mu\sigma}^\dagger d_{{\bf k}\nu\sigma}
\end{equation} 
in the plane-wave basis.
Here, the operator $d_{{\bf k} \mu \sigma}^\dagger$ 
($d_{{\bf k} \mu \sigma}$) creates 
(destroys) an electron with spin $\sigma$ and momentum 
${\bf k}$ in the $\mu$-th orbital. Matrix elements 
$\varepsilon^{\mu\nu}_{\bf k}$, which include 
both the hopping matrix elements and the 
momentum independent onsite orbital energies, 
are taken 
from Ref.[28]. The set of $d$-orbitals, to which 
orbitals $\mu$ and $\nu$ belong, consists 
of $d_{xz}$, $d_{yz}$, $d_{xy}$, $d_{x^2-y^2}$, and $d_{3z^2-r^2}$. 

The interaction part includes standard onsite
 Coulomb interactions
\begin{eqnarray}
{H}_{int} &=& U \sum_{{\bf i},\mu} n_{{\bf i}\mu \uparrow} 
n_{{\bf i}\mu \downarrow} + (U' -
\frac{J}{2}) \sum_{{\bf i}, \mu<\nu} n_{{\bf i} \mu} 
n_{{\bf i} \nu} \nonumber \\ 
&-& 2 J \sum_{{\bf i}, \mu<\nu} 
{\bf{S_{{\bf i} \mu}}}.{\bf{S_{{\bf i} \nu}}} + J \sum_{{\bf i}, \mu<\nu, \sigma} 
d_{{\bf i} \mu \sigma}^{\dagger}d_{{\bf i}
 \mu \bar{\sigma}}^{\dagger}d_{{\bf i} \nu \bar{\sigma}}
d_{{\bf i} \nu \sigma}. \nonumber\\
\label{int}
\end{eqnarray}
$U$ and $U^{\prime}$ are the intra-orbital and the 
inter-orbital Coulomb interaction, respectively. $J$ is the 
Hund's coupling, with the condition $U^{\prime}$
 = $U - 2J$ imposed for a rotation-invariant interaction. 

The mean-field Hamiltonian for the $(\pi, 0)$-SDW 
state in the two-sublattice 
basis is given by\cite{ghosh}
\begin{equation} 
\mathcal{H}_{mf} = \sum_{\bf k \sigma} 
\Psi^{\dagger}_{{\bf k} \sigma} (\hat{\zeta}_{{\bf k} \sigma} +
\hat{M}_{{\bf k} \sigma})
\Psi_{{\bf k} \sigma}.
\end{equation}
$\zeta_{{\bf k} \sigma}^{ll^{\prime}}$ are the 
matrix elements due to the kinetic part while 
$M^{ll^{\prime}}_{{\bf k} \sigma} = -s \sigma 
\Delta_{ll^{\prime}} \delta^{ll^{\prime}} + \frac{5J - 
U}{2} n_{ll^{\prime}} \delta^{ll^{\prime}}$. $l$,
 $l^{\prime}$ $\in$ $s \otimes \mu $ with $s$ and
$\mu$ belonging to the sublattice and orbital 
bases, respectively. Off-diagonal elements of 
$\Delta_{ll^{\prime}}$ 
and $n_{ll^{\prime}}$ are small for the parameters 
considered here, and hence neglected. $s$ and $\sigma$ in 
front of $\Delta_{ll^{\prime}} \delta^{ll^{\prime}}$ 
take value 1 (-1) for 
A (B) sublattice and $\uparrow$-spin 
($\downarrow$-spin), respectively. The electron
 field operator
is defined as $\Psi^{\dagger}_{\k \uparrow} = 
(d^{\dagger}_{A{\bf k}1 \uparrow},d^{\dagger}_{A{\bf k}2 \uparrow},.
.. ,d^{\dagger}_{B{\bf k}1 \uparrow},d^{\dagger}_{B{\bf k}2 
\uparrow},...)$, where subscript indices
1, 2, 3, 4, and 5 stand for orbitals $d_{3z^2-r^2}$, $d_{xz}$, 
$d_{yz}$, $d_{x^2-y^2}$, and $d_{xy}$, 
respectively. The exchange fields are given as $2\Delta_{ll} = 
Um_{l} + J \sum_{l \ne l^{\prime}}m_{l^{\prime}}$.
Orbital charge density and magnetization are determined 
in a self-consistent manner by diagonalizing the Hamiltonian. 

\begin{figure}[b]
\centerline{
\includegraphics[width=7.5cm,height=5.0cm,angle=0]{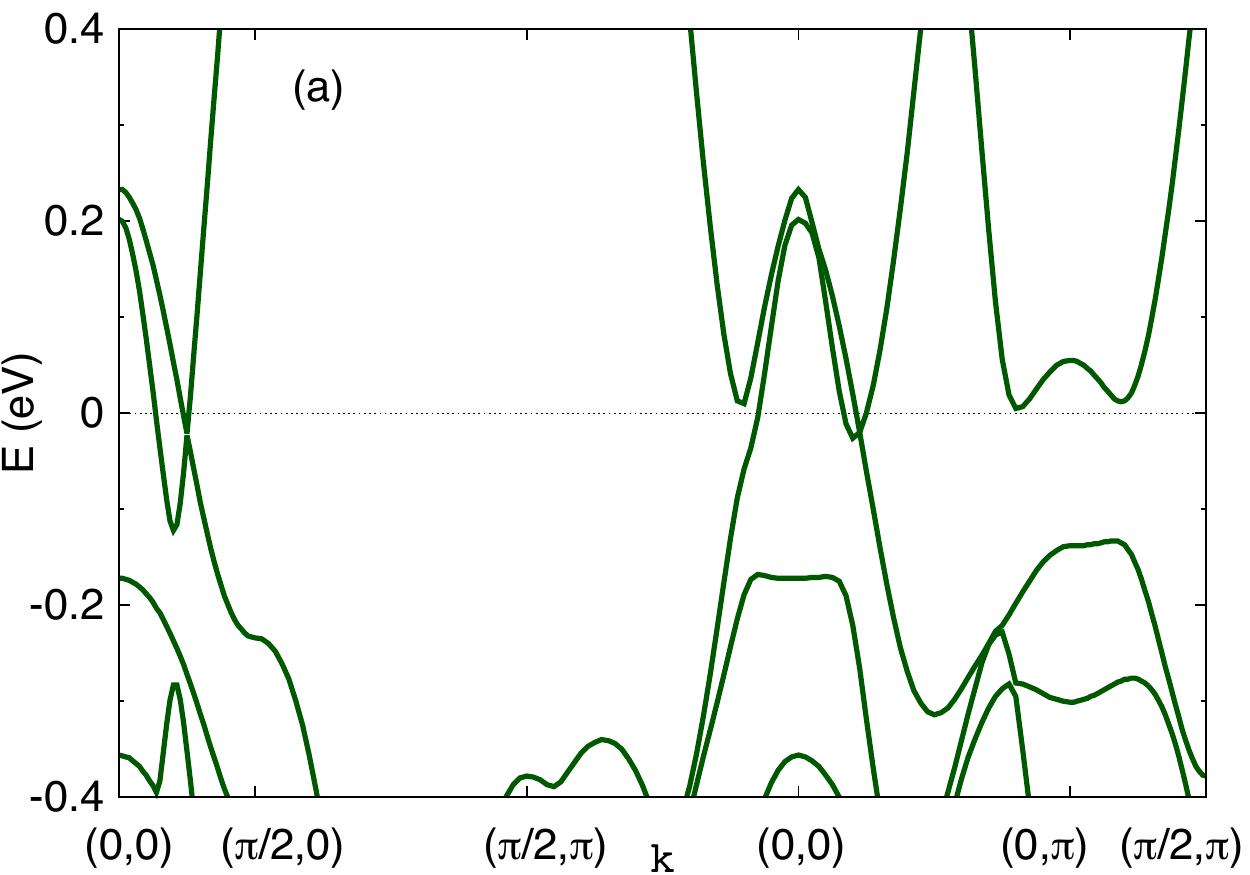}
}
\vspace{.2cm}
\centerline{
\includegraphics[width=8.0cm,height=3.8cm,angle=0]{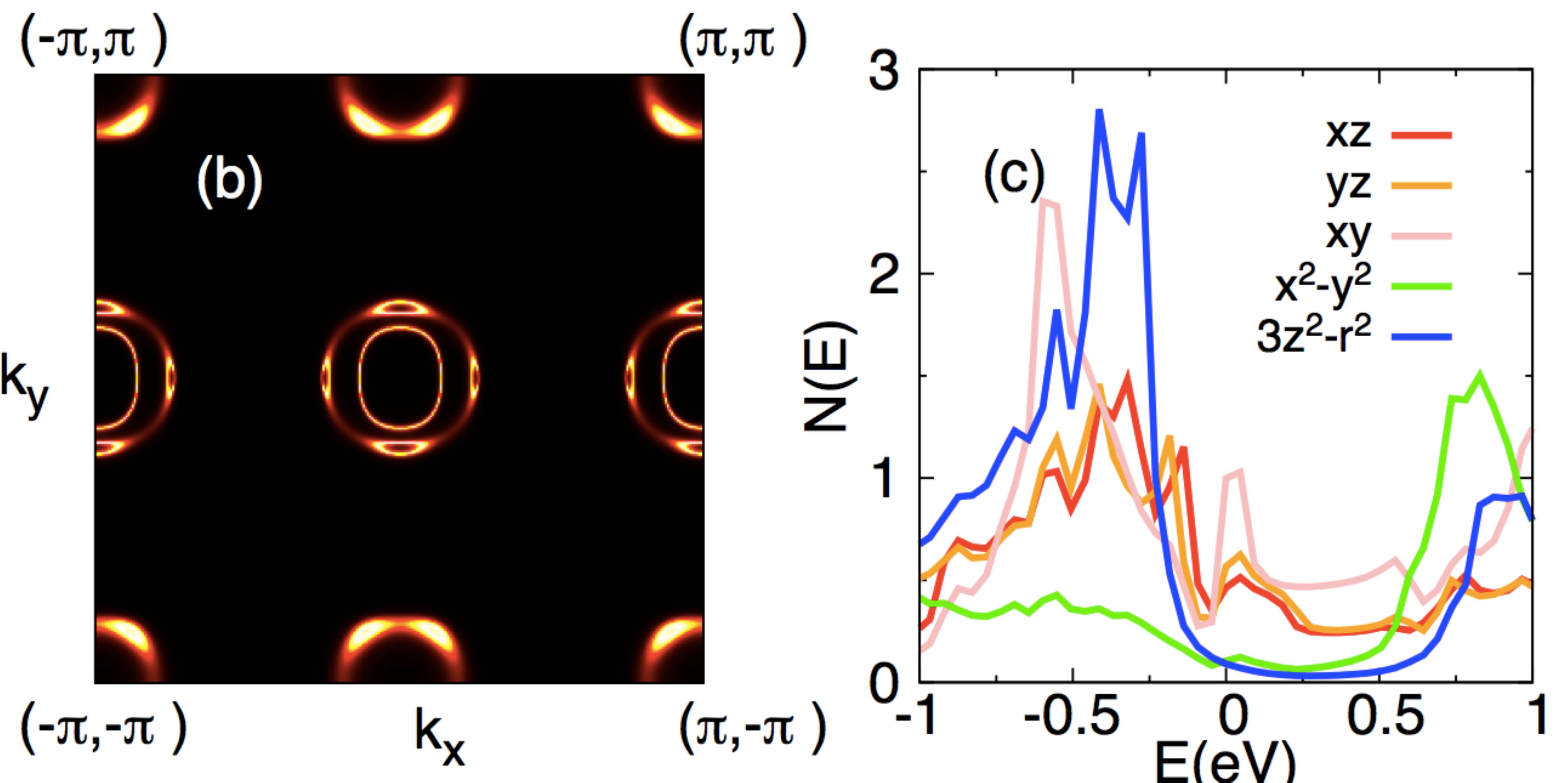}
}
\caption{(a)~Electronic dispersion along the high-symmetry directions, 
(b)~ Reconstructed Fermi surfaces 
consisting of several pockets near and around $\Gamma$ as well as around 
($\pi$, 0), and (c)~orbital-resolved density of states in the 
$(\pi, 0)$-SDW state.}
\label{disp}
\end{figure} 

The change caused in the Green's function because 
of a single impurity with $\delta$-potential 
is given by 
\be
\delta \hat{G}(\k, \k^{\prime}, \omega) = \hat{G}^0(\k,\omega)
 \hat{T}(\omega) \hat{G}^0(\k^{\prime},\omega)
\ee
using $t$-matrix approximation. $\hat{G}^0(\k,\omega) =
 (\hat{\bf I}- \hat{\mathcal{H}^{\prime}}_{mf})^{-1}$
is the Green's function in the SDW state with 
\be
\hat{\mathcal{H}}^{\prime \sigma}_{mf} = 
\begin{pmatrix}
 \hat{\varepsilon}_{\k} & {\rm sgn}\bar{\sigma}\hat{\Delta} \\
 {\rm sgn}\bar{\sigma}\hat{\Delta} & \hat{\varepsilon}_{\bf {k+Q}}
\end{pmatrix} .
\ee
$\hat{\bf I}$ is a 10$\times$10 identity matrix and ${\bf Q} 
= (\pi, 0)$. $\hat{\mathcal{H}^{\prime}}_{mf}$ is
obtained from $(\hat{\zeta}_{{\bf k}} +
\hat{M}_{{\bf k}})$ using a unitary transformation  \cite{dheeraj1} 
Next,
\be
T(\omega) = (\hat{\bf 1} - \hat{V}
 \hat{\mathcal{G}}^0(\omega))^{-1}\hat{V},
\ee
with 
\be
\hat{\mathcal{G}}^0(\omega) =
 \frac{1}{N} \sum_{\k} \hat{G}^{0}(\k, \omega)
\ee
and
\be
\hat{V} = V_{imp}
\begin{pmatrix} 
\hat{\bf 1} & \hat{\bf 1} \\
\hat{\bf 1} & \hat{\bf 1}
\end{pmatrix}.
\ee
Here, $\hat{\bf 1}$ is a 5$\times$5 identity matrix. 
The change $\delta \rho ({\bf q},\omega)$ in the 
DOS due to the impurity scattering is given by 
\be
\delta \rho({\bf q},\omega) = \frac{i}{2\pi} 
\sum_{\k} g(\k, {\bf q},\omega)
\ee
with
\be
g(\k, {\bf q},\omega) =  \text{Tr} \delta 
\hat{G}(\k,{\bf k^{\prime}},\omega)- 
\text{Tr} \delta \hat{G}^*({\bf k^{\prime}},\k,\omega),
\ee
where ${\bf k} - {\bf k}^{\prime} = {\bf q}$. 
The real-space QPI can be obtained as 
\be
\delta \rho({\bf r}_i,\omega) = \frac{1}{N} 
\sum_{\k} \delta \rho({\bf q},\omega) e^{i\k\cdot{\bf r}_i}.
\ee

In the following, intraorbital Coulomb interaction
 parameter ($U$) is taken as $1.07$eV 
with $J = 0.25U$ to keep the total magnetic 
moment per site less than unity. Band filling $n_e$ is fixed at 
6.03 (3$\%$ electron doping). 
Self-consistently obtained orbital magnetizations
are $m_{3r^2-x^2}  = 0.086$, $m_{xz} = 0.095$, $m_{yz} = 0.142$, 
$m_{xy} = 0.186$, and $m_{x^2-y^2} = 0.048$. Orbital 
charge densities are $n_{3r^2-x^2} = 1.469$, $n_{xz} = 1.208$,
$n_{yz} = 1.182$, $n_{xy} = 1.014$, and $n_{x^2-y^2} = 1.158$. 
Strength of the impurity potential $V_{imp}$
is set to be $200$meV. Varying strength will 
change the intensity while the basic structure of the QPI is 
expected to remain the same. A mesh size of 300 
$\times$ 300 in the momentum space is used
for all the calculations.

\begin{figure}[t]
\centerline{
\includegraphics[width=8.4cm,height=6.0cm,angle=0]{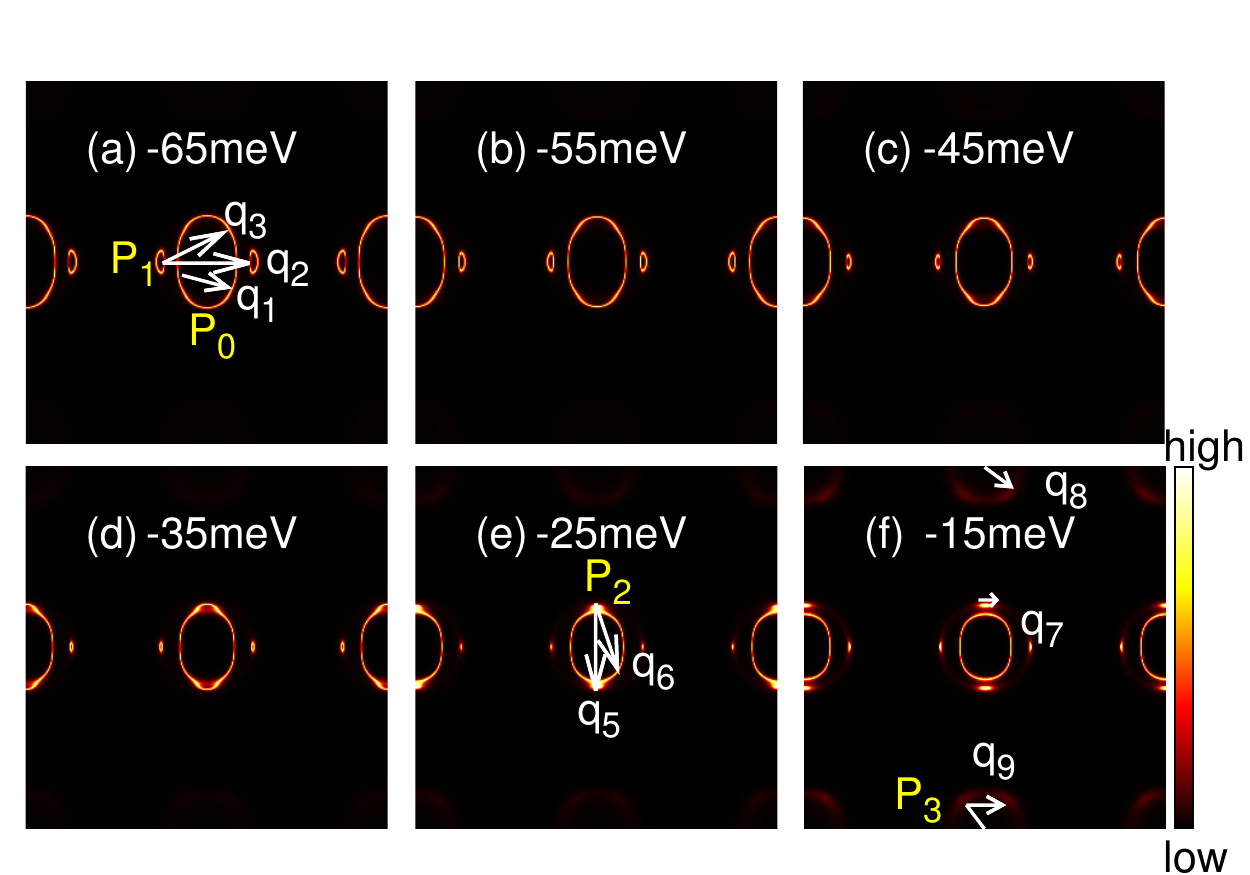}
}
\caption{Constant energy maps of the spectral 
functions $\mathcal{A}(\k, \omega)$ in 
the unfolded Brillouin zone from $-65$meV 
(top left) to $-15$meV 
(bottom right) in step of $10$meV.
The arrows represent scattering 
wavevectors in the SDW state. Note that ${\bf q}_4$, 
${\bf q}_7$, and 
${\bf q}_{10}$ are not shown, which are the
 intrapocket scattering vectors for the tiny CCEs 
P$_1$, P$_2$, and for the 
subpockets in CCE P$_3$, respectively.}
\label{spec}
\end{figure} 

\section{Results}

Fig.\ref{disp}(a) and (b) show the electronic 
dispersion and the Fermi surface (FS) 
in the SDW state. The FS 
consists of an ellipse-like hole pocket around 
$\Gamma$, with  major axis in the (0, 1) direction,
and tiny electron pockets situated at 
$\approx$ $(\pm \pi/4, 0)$ and $(0, \pm \pi/4)$ but 
outstretched along $(0, 1)$ and $(1, 0)$ directions,
 respectively. Interestingly, similar pockets although
larger in size have been reported by the ARPES 
experiments\cite{yi1,wang}. 
In addition, there are electron pockets around
($0, \pm \pi$).
 Some of the above characteristics 
of FSs lead to  significant anisotropy in the QPI.
Fig.\ref{disp}(c) shows that the $d_{xy}$, $d_{yz}$, and $d_{zx}$ 
orbitals dominate at the Fermi level.

\begin{figure}[b]
\centerline{
\includegraphics[width=8.4cm,height=6.0cm,angle=0]{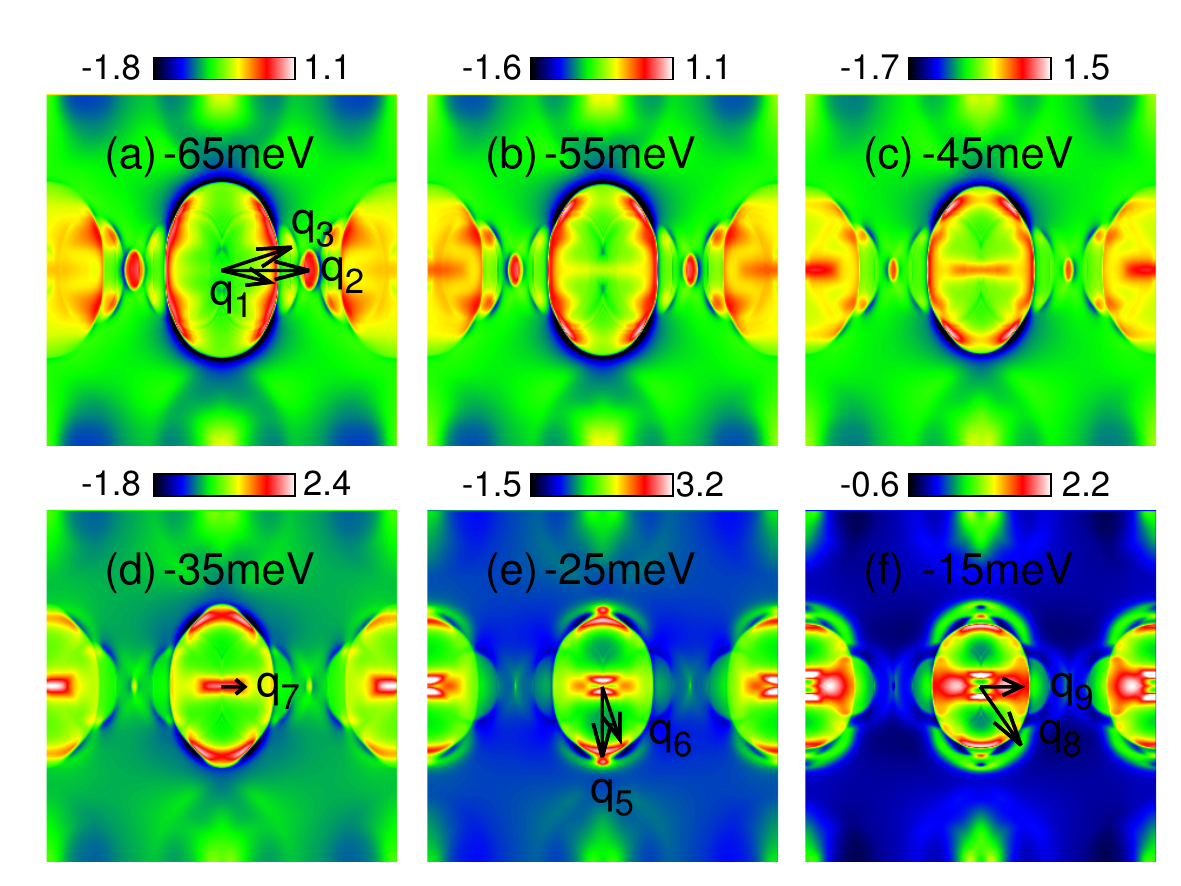}
}
\caption{QPI maps in the unfolded Brillouin zone for
different energies $\omega$ upto $-65$meV
(top left) to 
$-15$meV (bottom right) in steps of $10$meV.
The arrows denote
the QPI wavevectors in the SDW state.}
\label{qpim}
\end{figure}

In order to understand the QPIs, it will be useful
 to look at the CCEs of the spectral functions, 
which are shown in 
Fig.\ref{spec}(a)-(f) as a function of energy with 
step of $10$meV upto $-15$meV starting from 
$-65$meV. Near $-65$meV,
CCEs consists of an ellipse-like pocket P$_0$ around 
(0, 0) and two tiny pockets P$_1$ along $k_y = 0$ 
mapped onto each
other by 180$^{\circ}$ rotation owing to the $C_2$
 symmetry. Thus, there are four sets of 
scattering vectors -  
intrapocket scattering vectors ${\bf q}_1$ due to
 P$_0$, interpocket scattering vectors ${\bf q}_2$
 connecting 
the pockets of P$_1$, interpocket scattering 
vectors ${\bf q}_3$ connecting P$_0$ and
 P$_1$ and intrapocket
scattering vectors ${\bf q}_4$ (not shown in 
Fig.3(a)) due to P$_1$. Corresponding QPI 
pattern is expected to have 
a two-dimensional nature because of a near 
cancellation of two opposite tendencies 
in which scattering
vectors ${\bf q}_1$ tries to create a one-dimensional
 pattern along the $(0, 1)$ direction while ${\bf q}_2$ 
and ${\bf q}_3$ do the same along the $(1, 0)$
 direction. On the contrary, pattern consists 
of two parallel peak 
structures running along $q_x = const$ and 
passing through $q_x \approx$ ($\pm \pi/4, 0$) 
(Fig.\ref{qpim}(a)). 

\begin{figure*}
\begin{center}
\vspace*{-15mm}
\hspace*{0mm}
\psfig{figure=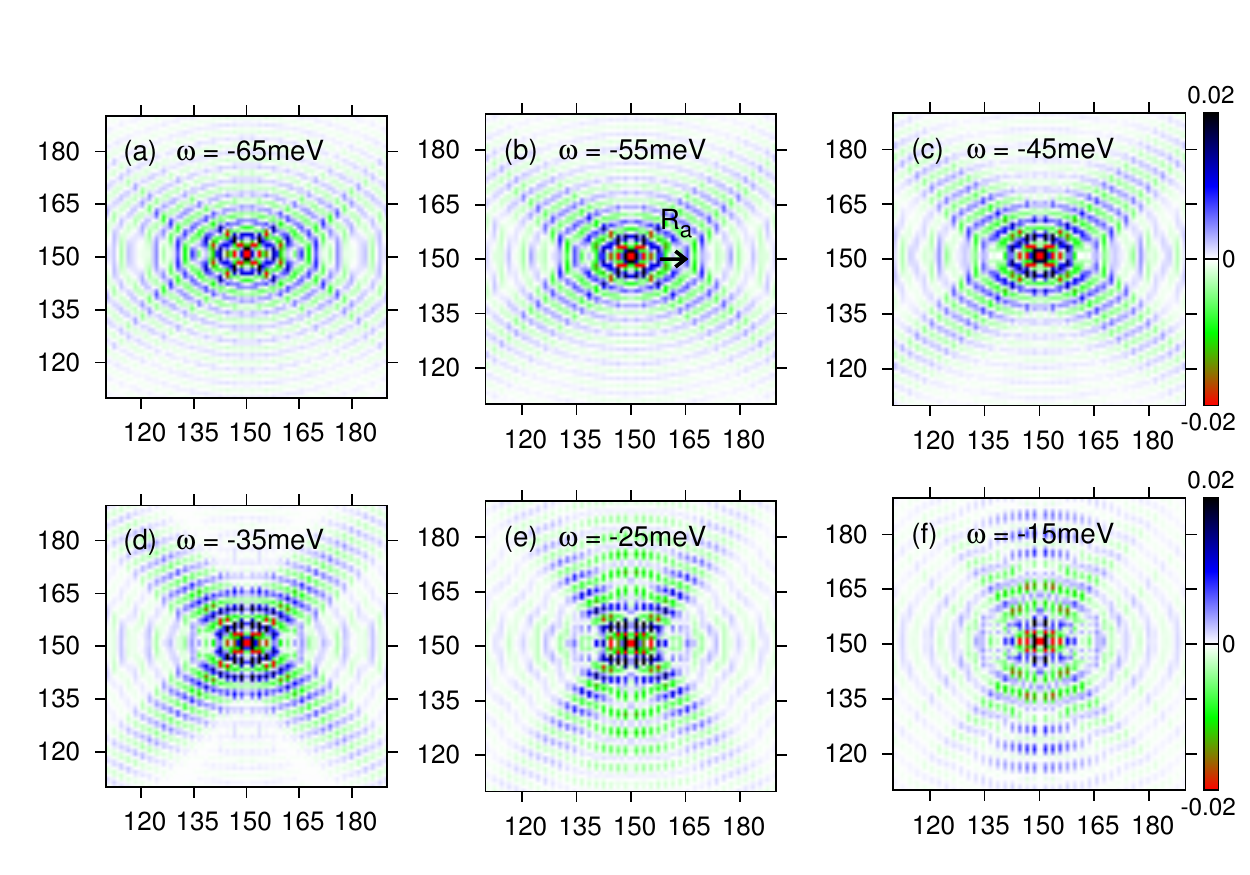,width=160mm,angle=0}
\vspace*{-9mm}
\end{center}
\caption{Real-space QPIs for the set of parameters as in
Fig.\ref{qpim}. LDOS modulation
along the antiferromagnetic direction ($x$-axis) with the
wavelength $R_x \approx 8a_{{\rm Fe}-{\rm Fe}}$. Although the 
period of modulation along the antiferromagnetic direction 
remains almost unchanged, the strong modulation direction is 
sensitive to the quasiparticle energy.}
\label{qpir}
\end{figure*}

In addition, there are small elliptical pockets
 located near $q_x \approx$ ($\pm \pi/2, 0$). 
Here, it is
important to note that only those scattering 
vectors are important, which connect parts 
of the CCEs having
same dominating orbitals because only 
intraorbital scattering has been incorporated 
owing to the symmetry consideration\cite{zhang}.
  
An important change in the QPI patterns occurs 
upon increasing the energy as shown in 
Fig.\ref{qpim}(d). This happens 
primarily because of the appearance of a new 
set of CCEs in the form of tiny pockets P$_2$, 
which emerge out of
the elliptical pocket P$_0$. Since P$_2$ is in the
 proximity of band extrema, pattern generated
 corresponding to 
the scattering vectors connecting these pockets 
should dominate the overall QPI pattern because
of a larger phase space available for the scattering
 processes. Therefore, the balance maintained by the 
two opposite tendencies described above is 
perturbed now. This results into a highly anisotropic QPIs
(Fig.\ref{qpim}(d)) dominated by wavevectors 
${\bf q}_5$, ${\bf q}_6$ and ${\bf q}_7$
 (Fig.\ref{qpim}(c)).
${\bf q}_5$ and ${\bf q}_7$ are the set of interpocket 
and intrapocket scattering vectors for P$_2$ pockets, 
respectively, whereas ${\bf q}_6$ is another set 
of interpocket scattering vectors connecting
 P$_0$ and P$_2$.

\begin{figure}[b]
\centerline{
\includegraphics[width=8.5cm,height=4.1cm,angle=0]{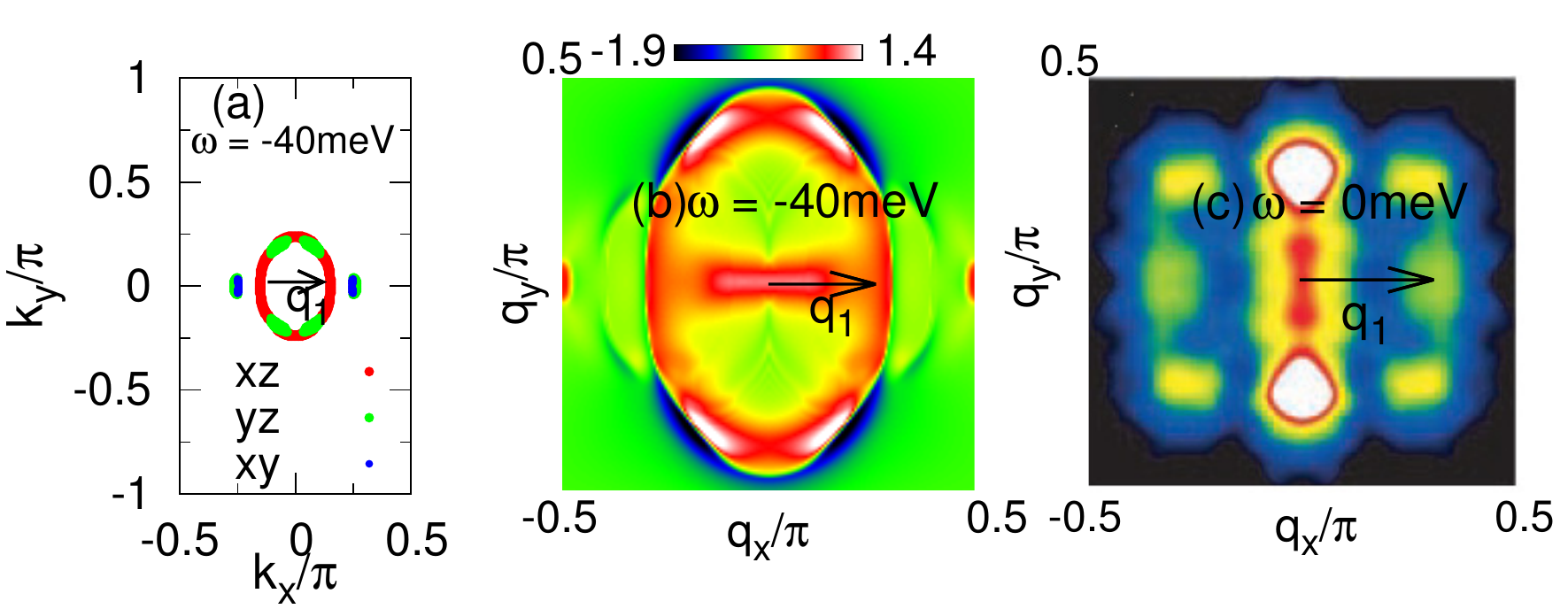}
}
\caption{(a)~CCEs for $\omega = $-40meV with
dominant orbital character in the SDW state
and (b)~corresponding
momentum-space QPI patterns. (c) QPI
patterns observed in
3$\%$ electron-doped CaFe$_2$As$_2$ for
$\omega = 0$meV by the STM.\cite{chuang}}
\label{qpi}
\end{figure}

On increasing energy further, an additional set of 
CCEs appear near ($0, \pm\pi$) as seen 
Fig.\ref{spec}(f), which may 
also contains very small subpockets (Fig.\ref{disp}(b)).
 As these are in the vicinity of local band extrema 
and parallel
to the $k_y = 0$, two-dimensional characteristics 
is imparted to the QPIs as noticed in Fig.3(f). 
Dominating QPI wavevectors 
are due to the inter subpocket scattering vectors 
${\bf q}_8$ and ${\bf q}_9$ as well as due to intra 
subpocket scattering vectors ${\bf q}_{10}$
 not shown (Fig.\ref{disp}(b)).  

Several aspects of the QPI obtained here compare 
well with those of SI-STM measurements carried out for
Ca(Fe$_{1-x}$Co$_{x}$)$_2$As$_2$. For $\omega 
= -35$meV, a central peak structure 
runs along $q_x = 0$ and consists of three main 
peaks which themselves
are made of multiple peaks either coinciding or 
placed very closely: one at (0, 0), and 
other two placed equidistant from it. Additionally, 
there are parallel running satellite peak 
structures situated at $\approx$ ($\pm \pi/4$, 0), 
which are part of an elliptical QPI patterns. 
The quasi-one dimensional nature of the pattern is 
found in a wide energy window. These features are 
in agreement with the STM measurements. We also note 
that they very sensitive to energy as evident 
from Fig.\ref{qpim}(e), when they become very weak
 for $\omega  = -15$meV
on increasing the quasiparticle energy further. 
Features of the 
CCEs especially the existence of pockets along 
$k_x = 0$, which play a crucial role in imparting 
the quasi-one dimensionality to the main peak 
structure, have also been noticed in 
the ARPES measurements. We have also examined 
the role of an orbital splitting term in the
Hamiltonian, which is found to bring in only minor 
deviations because of a relatively 
large sized P$_1$ and also  due to the suppression 
of pockets P$_2$. Thus, it is the significant 
band reconstruction in the SDW state which is 
responsible for the experimentally observed anisotropy in the 
QPI.

Fig.\ref{qpir} shows corresponding real-space QPI
 pattern. The wavelength of LDOS modulation 
along $x$ (antiferromagnetic direction) is 
$R_x \sim 8a_{{\rm Fe}-{\rm Fe}}$ for all 
energy values considered here 
though modulation may be weak or strong 
depending on the energy. A strong modulation 
along $x$ is seen for energies $\omega = -65$meV
{\rm and} $-55$meV, with 
parallel running peak structures along
 $q_x = const$ and 
passing through $q_x = 2\pi/R_x \approx$ ($\pm \pi/4, 0$) 
in qualitative agreement with the impurity 
induced electronic structure observed by the 
SI-STM experiments.
However, it becomes stronger along $x \approx 
y$ upon increasing the energy and 
corresponds to a strong modulation of DOS in
 the momentum space along a direction tilted 
away from  $q_x = const$. 

\section{Discussion} 
 
\subsection{Physical mechanism}

Our study highlights the role of redistributed 
orbital weight along the CCEs. In the unordered state, 
the hole pockets around $\Gamma$ have predominantly $d_{xz}$ and 
$d_{yz}$ character distributed in way 
to respect the four-fold rotational symmetry. On the
 contrary, the CCE pocket around $\Gamma$ in 
the SDW state is dominated by the
$d_{xz}$ orbital. For $\omega = -65$meV, the 
scattering vectors connecting the regions near 
the vertices along the minor axis 
of elliptical pocket leads to the most intense region 
in the QPI. This happens primarily because of two reasons.
First, a larger phase space is available when compared with 
the case of scattering vectors connecting the vertices along the 
major axis. Secondly, only intraorbital scattering processes
are taken into account. On increasing energy, a small $d_{yz}$ rich region
 (Fig.\ref{qpi}(a)) appears along the elliptical
CCE as a new band crosses the quasiparticle energy.
 Because of the new band's extrema and associated large spectral weight, 
QPI pattern due to the scattering vectors 
connecting $d_{yz}$ rich regions prevails over the 
others (Fig.\ref{qpi}(b)). Consequently, the most 
intense region moves towards the vertices along the major axis 
(Fig.\ref{qpim}(b)-(d)) a features present also in
the STM results (Fig.\ref{qpi}(c)).

In this work, the focus was on the LDOS modulation.
Another important issue is the modulation in 
the local magnetization induced by in the vicinity of the 
impurity.  This has been addressed in a recent work within 
a self-consistent approach for single 
non-magnetic impurity\cite{gastiasoro}.
The study found that the impurity induces magnetic nanostructures
with checkerboard-type order inside,  extended along
the antiferromagnetic direction with a significant LDOS 
modulation at the ends. Our result on the LDOS modulation 
in real-space is also consistent with this  study.

\subsection{Comparison with earlier work}

Anisotropy in the QPI patterns of the SDW state is not
unexpected because of the breaking of four-fold rotational 
symmetry. However, the details of the patterns depend on the electronic 
structure. A plausible description of the QPI patterns restrict the 
modeling of the electronic structure. The failure  
of almost all the earlier work in reproducing the 
nearly parallel
running satellite peak structures along $q_x \approx \pm\pi$/4 
highlights  the limitation of the electronic structure used.
In our work, these structures result from the elliptical CCE
around $(0, 0)$ with a major role played by the scattering vectors 
lying nearly parallel to the minor axis of length $\sim \pi$/4. Thus, we 
believe that the Fermi pocket around $\Gamma$, the
existence of which has 
also been suggested by the ARPES measurements,
is likely to be elliptical in shape.

\subsection{Unresolved issues}

For 3$\%$ doping on the parent state the 
$8a_{{\rm Fe}-{\rm Fe}} \times 8a_{{\rm Fe}-{\rm Fe}}$
nanostructures would contain more than one impurity atom 
on the average.
Therefore, the interference between scattering events from 
multiple impurities 
could be important for the measured QPI patterns. The 
present t-matrix approach unfortunately does not access
these effects. Recently, a framework to study 
QPI in the presence of interacting multiple 
impurities has been discussed\cite{mitchell}.
However, in many instances, single
impurity treatment has yielded QPI patterns which 
successfully describe the qualitative features of
STM measurements. The LDOS
modulation obtained in this work with the
periodicity $\sim 8a_{{\rm Fe}-{\rm Fe}}$
is another such example.

LDOS modulation with the experimentally 
observed periodicity is reproduced successfully in our results 
along the antiferromagnetic direction, 
and is robust against any change in the quasiparticle energy.
 However, the strongly modulated direction exhibits 
 sensitivity to the quasiparticle energy, which is 
 due to the fast change in the CCEs. In the 
 experiments, however, 
the strongly modulated direction is robustly along the
 antiferromagnetic direction despite the change in energy. This may
 indicate that CCEs change comparatively slowly in the real systems
 as a function of energy.

\section{Conclusions}

We have investigated the quasiparticle interference pattern
in the ($\pi, 0$)-SDW state
using a five-orbital tight-binding model of electron-doped 
iron pnictides.  With a realistic reconstructed bandstructure,
which includes an ellipse-like constant energy
contour around (0, 0) and additional nearby smaller pockets,
we find highly anisotropic QPI patterns. Because the 
scattering vectors oriented along the minor axis of the 
elliptical CCE (of length $\pi/4$) connects $d_{xz}$ rich segments,
QPI peak structures are obtained 
at $\approx$ ($\pm \pi/4, 0$), running parallel to the
$q_y$ axis.
The corresponding real-space pattern consists of LDOS modulation 
along the antiferromagnetic direction with periodicity
$\sim$ $8a_{{\rm Fe}-{\rm Fe}}$. Both the features are in 
agreement with STM results for the doped iron pnictides.

We acknowledge use of the HPC Clusters at HRI.

\end{document}